\documentclass[aps,pra,twocolumn,nofootinbib,citeautoscript,10pt]{revtex4-1}  
\usepackage{amsmath,amssymb,bm, makecell} 

\usepackage{xcolor}
\usepackage{svg}

\usepackage{graphicx}

\usepackage{color} 
\usepackage[papersize={8.5in,11in}]{geometry}
\usepackage[colorlinks=true]{hyperref}
\hypersetup{        
    unicode=false,          
    pdftoolbar=true,        
    pdfmenubar=true,        
    pdffitwindow=false,     
    pdfstartview={FitH},    
    pdfkeywords={keyword1} {key2} {key3}, 
    pdfnewwindow=true,      
    colorlinks=true,       
    linkcolor=magenta, 
    citecolor=blue,        
    filecolor=magenta,      
    urlcolor=blue           
} 

\usepackage{graphicx}
\usepackage{dcolumn}
\usepackage{color}
\usepackage{amssymb,amsmath}
\usepackage{tabularx,graphicx}
\usepackage{epstopdf}
\usepackage{latexsym}
\usepackage{colortbl}
\usepackage{psfrag}
\usepackage{bbm,bm,array,physics}
\usepackage{dsfont}

 \def\*#1{\mathbf{#1}} 

\newcommand{\be}{\mathbf{e}}

\def\be{\begin{eqnarray}}
\def\ee{\end{eqnarray}}
\def \be{\begin{align}}
\def \ee{\end{align}}
\def \bea{\begin{eqnarray}}
\def \eea{\end{eqnarray}}

\begin{document}

\title{Strong pair-density-wave fluctuations in an exactly solvable doped Mott insulator}

\author{Igor de M. Froldi}
\affiliation{Instituto de F{\'i}sica, Universidade Federal de Goi{\'a}s, 74.690-900,
Goi{\^a}nia-GO, Brazil}

\author{Carlos Eduardo S. P. Corsino}
\affiliation{Instituto de F{\'i}sica, Universidade Federal de Goi{\'a}s, 74.690-900,
Goi{\^a}nia-GO, Brazil}

\author{Hermann Freire}
\affiliation{Instituto de F{\'i}sica, Universidade Federal de Goi{\'a}s, 74.690-900,
Goi{\^a}nia-GO, Brazil}

\begin{abstract}
We investigate the Hatsugai-Kohmoto (HK) model on a square lattice, which describes both a Mott insulator at half-filling and a non-Fermi liquid phase on doping. Through the solution of this exactly solvable model with the inclusion of pairing interactions, we demonstrate the emergence of strong pair-density-wave (PDW) fluctuations associated with center-of-mass momentum {$\mathbf{Q}=(\pi,\pi)$} at finite temperatures for low dopings up to a critical doping and intermediate $U$ interaction. Furthermore, we {also} confirm that a superconducting instability  appears in the model {within a wide regime of} interaction $U$ and doping parameter $x$.
In view of the fact that it has been recently put forward that the metal-insulator transition of the HK model belongs to the same universality class as the Mott transition of the paradigmatic Hubbard model [Huang \emph{et al.}, Nat.
Phys. \textbf{18}, 511 (2022)], our work may thus shed light on an interesting scenario regarding the emergence of a fluctuating PDW phase on doping a Mott insulator, which has been argued to be relevant for understanding the physics of the cuprate superconductors in the underdoped regime. 
\end{abstract}

\maketitle

\section{Introduction} 

The construction of a robust microscopic theory explaining the mechanism of a pair-density-wave (PDW) phase \cite{Agterberg_2020}, i.e., an unconventional superconducting state with finite Cooper-pair center-of-mass momentum with no explicit breaking of time-reversal symmetry, remains a formidable challenge to this date. The underlying motivation for this investigation is that, from an experimental point of view, a PDW phase has been recently detected in a number of correlated materials, which include, e.g., Kagome metals \cite{Chen_2021,Ge_2024}, the UTe$_2$ compound \cite{Jiao_2020,Gu_2023}, and some Fe-based superconductors \cite{Liu_2023,Zhao_2023}. Another reason for the widespread interest in this correlated phase is due to the fact that the possible existence of fluctuations associated with such an order has been argued \cite{PALee_2014} to be relevant for understanding some aspects
of the physics of cuprate superconductors in the underdoped regime, such as, e.g., LBCO at 1/8 hole doping and some compounds in the La-214 family \cite{Zhang_Tranquada,Tranquada_2007,Berg_2009,Lozano_2023} and in the Bi$_2$Sr$_2$CaCu$_2$O$_{8+\delta}$ (Bi-2212) \cite{Hamidian_2016,Du_2020}.

On the theoretical side, the existence of a PDW phase has only been established in a few quasi-one-dimensional models, such as, {e.g.}, the Heisenberg antiferromagnetic chain interacting via a Kondo coupling to a one-dimensional Luttinger liquid \cite{Berg_2010}, among other examples (see, {e.g.}, Refs. \cite{Jiang_2023,Ashvin_2022}), using the density matrix renormalization group (DMRG) approach. However, extending this analysis to fully two-dimensional models has proven to be very difficult and, for this reason, progress in this direction has been somewhat hindered. In recent years, many additional works have appeared employing a wide variety of techniques, all of which reporting evidence on the existence of such an unconventional superconducting phase in two dimensions, both approximately in some correlated models (see, {e.g.}, Refs. \cite{Berg2_2009,Loder_2010,Soto_2014,Wang_2015,Freire_2015,Wardh_2017,Carvalho_2021,Wu_2023,Santos_2023,Setty2_2023}) or exactly in other models \cite{Coleman_2022,Setty_2023} and also numerically \cite{Corboz_2014,Jiang_Kivelson,Xiao_Lee,jiang2023pair,liu2024enhanced,zhu2024exact,wang2024pair}. These works suggest that  intermediate-to-strong interactions may be needed in order to unravel the microscopic mechanism underlying the formation of PDW phases in these systems.

In the present work, we address this fundamental problem from the perspective of an exactly solvable strongly correlated model in two dimensions -- the Hatsugai-Kohmoto (HK) model \cite{HK_model} (see also Ref. \cite{Baskaran} for a similar model). The HK model was proposed in the early 1990s as a minimal model for describing both a Mott insulating state \cite{HK_model,Continentino,M_Coutinho,Yeo_Phillips} at half-filling and a non-Fermi liquid (NFL) phase that emerges on doping \cite{Phillips2020}. 
It involves a kinetic term with nearest-neighbor hopping amplitude $t$ on a square lattice and an all-to-all constant interaction $U$ among the fermions\footnote{It is worth emphasizing that the all-to-all interaction in the HK model is constant, rather than random. Therefore, the HK model should be contrasted with the well-known Sachdev-Ye-Kitaev (SYK) model \cite{SYK,SYK_review}, which in turn invokes the presence of disorder as a key ingredient for its exact solution. Interestingly, it was shown in Ref. \cite{Setty_2020} that the HK model within the Mott insulating regime does exhibit SYK dynamics.}, which permits the model to be analytically tractable.
For this reason, the HK model can be viewed as a simplified system that arguably has some analogies, e.g., to the paradigmatic and hard-to-solve two-dimensional Hubbard model. The latter is generally assumed to be central for capturing the physics of many unconventional high-$T_c$ superconductors \cite{PALee_review}. In the past few years, there
has been an increase of interest in the investigation of the HK model with the derivation of many new results, some of which consisting of, {e.g.}, the demonstration of a superconducting (SC) instability from the NFL phase  \cite{Phillips2020,Li2022,Zhu_2021}, the existence of Fermi arcs and a pseudogap if the corresponding interaction becomes momentum dependent \cite{Yang_HK}, and the exact computation of many thermodynamic properties \cite{Zhao_HK,Yeo_Phillips} and also of topological properties (see, {e.g.}, Refs. \cite{Zhu_Wang,BBradlyn,Zhao_Phillips,Setty_Si,Mai_Phillips,Byczuk,Marcin_2023}).

In addition to the fact that the HK model hosts a Mott insulator at zero doping and a NFL state with an instability towards a superconducting phase on doping, the structure of the single-particle Green's function of the model at half-filling exhibits some similarities with a phenomenological theory that describes the properties of the pseudogap state in underdoped cuprates: the so-called Yang-Rice-Zhang (YRZ) ansatz \cite{YRZ_ansatz}. Furthermore, it has been recently argued that the metal-insulator transition of the HK model lies in the same universality class of the Mott transition in the Hubbard model \cite{Huang_FixedPoint}. As a result, it seems reasonable to consider the HK model as a possible starting point that may also give some insight into other (potentially universal) aspects of the physics of strongly correlated materials. Given the conjectured relevance of the scenario of PDW fluctuations \cite{Agterberg_2020,PALee_2014} as either a ``mother state'' or a competing phase for describing the cuprate compounds in the underdoped regime, one important question that we wish to address in this work is whether the HK model also exhibits such an order, at least in some part of its phase diagram. As will become clear shortly, we will answer this question affirmatively. We will then explore an interesting scenario that arises as a consequence of this result, which could emerge in real materials.

We organize the remainder of this paper as follows: In Sec. \ref{HKmodel}, we define the HK model and discuss some previous exact results obtained for this system that, in fact, were part of the motivation of the present study. In Secs. \ref{PDW} and \ref{Tc}, we set out to show that strong PDW fluctuations do emerge in the model for low dopings up to a critical doping regime for intermediate interactions, and then we discuss some potential consequences of this result. Finally, in Sec. \ref{Conclusions}, we end with our conclusions and also provide an outlook of some future directions of our present work.

\section{Hatsugai-Kohmoto model}
\label{HKmodel}

As mentioned before, our starting point is the Hamiltonian of the strongly correlated HK model on a square lattice \cite{HK_model,Phillips2020}, which is given by 
\begin{align}\label{PDWHamiltonian}
    H_{HK}  &= \sum_{\mathbf{k}, \sigma} \left( \epsilon_\mathbf{k} - \mu \right)c^{\dagger}_{\mathbf{k} \sigma}c_{\mathbf{k} \sigma} + U \sum_{\mathbf{k}} n_{\mathbf{k} \uparrow} n_{\mathbf{k} \downarrow},
\end{align}
where $c^{\dagger}_{\mathbf{k} \sigma}$ and $c_{\mathbf{k} \sigma}$ are, respectively, the creation and annihilation of electrons with momentum $\mathbf{k}$ and spin projection $\sigma=\uparrow,\downarrow$, $\epsilon_\mathbf{k}=-2t(\cos k_x +\cos k_y)$ is the band dispersion with $t$ being the hopping parameter, $\mu$ is the chemical potential, $n_{\mathbf{k} \sigma}=c^{\dagger}_{\mathbf{k} \sigma}c_{\mathbf{k} \sigma}$ is the number operator and $U>0$ represents a local repulsion in momentum space. Because of the latter choice of interaction, the momentum $\mathbf{k}$ clearly remains a good quantum number. Because of this, since all eigenstates turn out to have a fixed nonfluctuating occupancy $n_{\mathbf{k} \sigma}$ in momentum space, the ground state factorizes as a tensor product of such states at different momentum and the model becomes exactly solvable. 

One interesting aspect of this model is that, even though the kinetic and potential energy operators commute, as can be seen in Eq. \eqref{PDWHamiltonian}, it nevertheless displays a metal-insulator transition induced by correlations as a function of $U/t$. Therefore, this Hamiltonian emerges as a minimal model in order to understand (exactly) how the proximity of a Mott insulator affects the strongly correlated non-Fermi liquid and superconducting phases described by it.

Due to the repulsive interaction $U$, the momentum space in the ground state becomes divided into three different regions: empty region ($\Omega_0$), single-occupied region ($\Omega_1$), and double-occupied region ($\Omega_2$). As a result, the ground-state expectation value of the number operator, $\bra{g}n_{\mathbf{k} \sigma} \ket{g} \equiv \langle n_{\mathbf{k} \sigma} \rangle$ (where $\ket{g} = \left[\Pi_{\mathbf{k} \in \Omega_2} c^{\dagger}_{\mathbf{k} \uparrow} c^{\dagger}_{\mathbf{k} \downarrow} \right] \left[ \Pi_{\mathbf{k} \in \Omega_1} \frac{1}{\sqrt{2}} (c^{\dagger}_{\mathbf{k} \uparrow} + c^{\dagger}_{\mathbf{k} \downarrow}) \right] \ket{0}$) at zero temperature becomes\footnote{We assume here that the $\Omega_1$ region in the ground state of the HK model is not spin polarized \cite{Li2022}.}:
\begin{align} \label{number1}
    \langle{n_{\mathbf{k} \sigma}}\rangle = 
        \begin{cases}
            0, & \mu < \epsilon_\mathbf{k} < \frac{W}{2} \hspace{1.5cm}(\mathbf{k}\in\Omega_0),\\
            \frac{1}{2}, &  \mu - U < \epsilon_\mathbf{k} < \mu \hspace{0,97 cm}(\mathbf{k}\in\Omega_1),\\
            1, & - \frac{W}{2} < \epsilon_\mathbf{k} < \mu-U \hspace{0.54cm}(\mathbf{k}\in\Omega_2),
        \end{cases}
\end{align}
where $W=8t$ is the bandwidth. We point out that, due to the existence of a single-occupied region $\Omega_1$, the ground state of the model displays a huge degeneracy, which will be lifted due to a small perturbation that will be added shortly to the system. Writing Eq. \eqref{number1} in a more compact way, we have:
\begin{align}\label{NumberOp}
    \langle n_{\mathbf{k}\sigma} \rangle = \frac{1}{2} \Theta\left( \epsilon_\mathbf{k} - \mu + U \right) \Theta \left( -\epsilon_\mathbf{k} + \mu \right) + \Theta\left( -\epsilon_\mathbf{k} + \mu -U  \right).
\end{align}
Hence, it can be shown that the exact retarded Green's function of the HK model is given by \cite{Phillips2020}:
\begin{align}\label{ExactGF}
G_{\sigma}(\mathbf{k},i\omega)=\frac{1- \langle{n_{\mathbf{k} \bar{\sigma}}}\rangle}{i\omega-\xi_{\mathbf{k}}}+\frac{\langle{n_{\mathbf{k} \bar{\sigma}}}\rangle}{i\omega-\xi_{\mathbf{k}}-U},
\end{align}
where $\xi_{\mathbf{k}} \equiv\epsilon_{\mathbf{k}}-\mu$ and $\bar{\sigma}=-\sigma$. Therefore, the low-energy effective description of the HK model possesses two ``emergent Fermi surfaces'' (associated with the lower and upper Hubbard bands) with their corresponding fractionalized excitations resembling the holon and doublon composite operators generally used as an approximation to describe the Hubbard model near the atomic limit \cite{Roth_1969,Haurie_2024}.

\begin{figure}[t]
\centering
\includegraphics[width=1\linewidth]{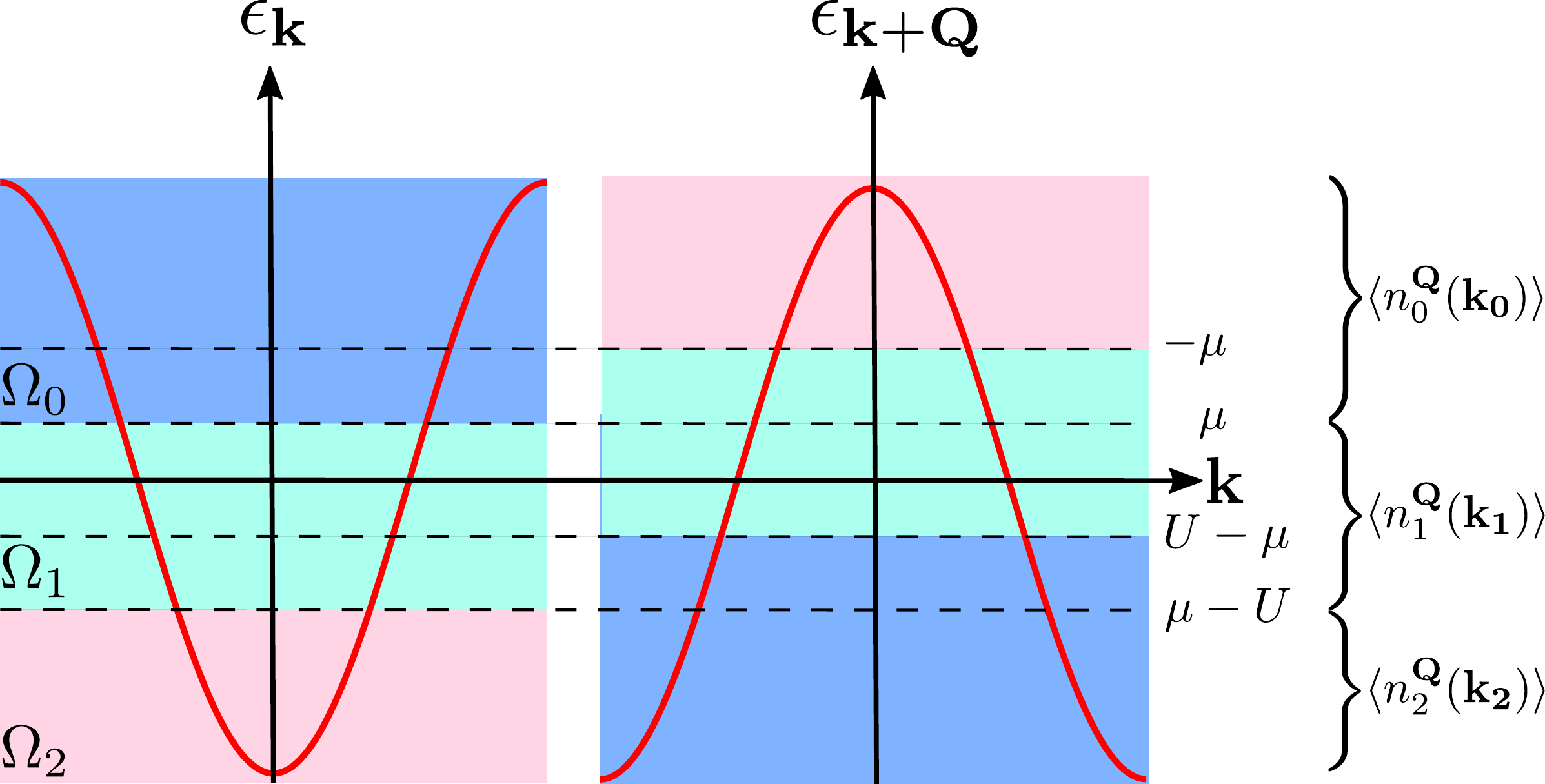}
\caption{Schematic representation of the band dispersions $\epsilon_{\mathbf{k}}$ and $\epsilon_{\mathbf{k+Q}}$ [for $\mathbf{Q}=(\pi,\pi)$] as a function of $\mathbf{k}$ in the HK model. Note the overlap of the momentum regions for this particular case of $\mu$ and $U$: $\Omega_0$ on the left (dark blue) translates into $\Omega_2$ and $\Omega_1$ regions on the right, while $\Omega_1$ on the left (light blue) translates into itself and $\Omega_0$ on the right, and lastly $\Omega_2$ on the left (light red) translates into $\Omega_0$ on the right. These are indicated by the curly brackets on the right with $n^{\mathbf{Q}}_i(\mathbf{k_i})$ ($i=$ 0, 1 and 2), given by Eqs. \eqref{n0} to \eqref{n2}.}
\label{Overlap}
\end{figure}

Since $\langle{n_{\mathbf{k} \sigma}}\rangle= \langle{n_{\mathbf{k} \bar\sigma}}\rangle$ due to
spin-rotation invariance, $G_{\sigma}(\mathbf{k},\omega)$ actually does not depend on the spin projection. As for the doping parameter $x$ of this model, which is given by $x=1-\sum_{\mathbf{k}\sigma} \langle n_{\mathbf{k}\sigma} \rangle$ (the ``volume'' of the system has been set equal to unity), we get by demanding that the $\Omega_1$ region lies within the band the following conditions at $T=0$: 
\begin{align}\label{ZeroTempDoping}
x=\begin{cases}
    \frac{1}{2}- \tilde{\mu}, \mbox{ if } \tilde{\mu} - {u} < -\frac{1}{2} \mbox{ and } -\frac{1}{2}<\tilde{\mu} \leq \frac{1}{2},\\
    {u} - 2\tilde{\mu}, \mbox{ if } \tilde{\mu} - {u} > -\frac{1}{2} \mbox{ and } -\frac{1}{2}<\tilde{\mu} \leq \frac{1}{2},\\
    {u} - \tilde{\mu} - \frac{1}{2}, \mbox{ if } -\frac{1}{2} \leq \tilde{\mu} - {u} \leq \frac{1}{2} \mbox{ and } \tilde{\mu} \geq \frac{1}{2},\\
    0, \mbox{ if } \tilde{\mu} - {u} < -\frac{1}{2} \mbox{ and } \tilde{\mu} > \frac{1}{2},
\end{cases}
\end{align}
where ${u}={U}/{W}$ and $\tilde{\mu}={\mu}/{W}$. At finite temperatures, since we must fix the overall charge density in the HK model given by:
\begin{align}
2\int\frac{d^2\mathbf{k}}{(2\pi)^2}\frac{e^{-\beta\xi_{\mathbf{k}}}+e^{-\beta(2\xi_{\mathbf{k}}+U)}}{1+2e^{-\beta\xi_{\mathbf{k}}}+e^{-\beta(2\xi_{\mathbf{k}}+U)}}=1-x,
\end{align}
by solving the above equation, we obtain that, for a fixed $x$ and $U$, $\mu$ varies as a function of $T$. In Appendix \ref{Appendix_A}, we show some expressions of $\mu(T,x,U)$ for physically interesting regimes.

For the case of half-filling $x=0$ and $U>W$,  it can be shown that the exact Green's function given in Eq. \eqref{ExactGF} becomes:
\begin{align}
G_{\sigma}(\mathbf{k},i\omega)=\frac{1}{i\omega-(\xi_{\mathbf{k}}+U/2)-\left(\frac{U^2}{4}\right)\frac{1}{i\omega-(\xi_{\mathbf{k}}+U/2)}}.
\end{align}
Since both real and imaginary parts of the self-energy diverge at $\omega=0$ for $\xi_{\mathbf{k}}=-U/2$, this quantity turns out to be clearly nonanalytical. As a result, the latter property violates the central tenet that forms the basis of the effective Landau's Fermi liquid theory and indicates that the HK model indeed describes a non-Fermi liquid.

\section{Pair binding energies for PDW and SC in the HK model} \label{PDW}

Following a straightforward generalization of Cooper's original argument \cite{Cooper_1956,BCS_1957}, we assume that there is an infinitesimal attraction in a pairing channel associated with momentum $\mathbf{q}$. This leads to the definition of a HK-SC model described by $H   = H_{HK} + H_{V}$, where:
\begin{align}\label{PDWHamiltonian2}
    H_{V} &=  - V_{\mathbf{q}}\sum_{\mathbf{k,k'}} c^{\dagger}_{\mathbf{k + q} \uparrow} c^{\dagger}_{\mathbf{-k} \downarrow} c_{\mathbf{-k'} \downarrow} c_{\mathbf{k' + q} \uparrow}.
\end{align}
For the analysis of the instability towards a PDW phase, the ground-state expectation value of the operator $n_{\mathbf{k+q},\sigma}$ can be calculated from Eq. \eqref{NumberOp}. In what follows, we will use the notation where the subscript $i$ in the momentum $\mathbf{k_i}$ refers to the momentum region, {i.e.}, $\mathbf{k_i} \in \Omega_i$ (for $i=0, 1, 2$), and $n^{\mathbf{q}}_i(\mathbf{k_i}) \equiv  \langle n_{\mathbf{k_i+q},\sigma} \rangle$. Therefore, we can write:
\begin{align}\label{n0}
    n^\mathbf{q}_0 (\mathbf{k_0}) = &  \mathcal{F}(\mathbf{k_0 + q}) \Theta \left( \epsilon_{\mathbf{k_0}} - \mu \right),\\\label{n1}
    n^\mathbf{q}_1 (\mathbf{k_1}) = &  \mathcal{F}(\mathbf{k_1 + q}) \Theta \left( -\epsilon_{\mathbf{k_1}} + \mu \right) \Theta \left( \epsilon_{\mathbf{k_1}} - \mu + U \right), \\\label{n2}
    n^\mathbf{q}_2 (\mathbf{k_2}) = & \mathcal{F}(\mathbf{k_2 + q}) \Theta \left( -\epsilon_{\mathbf{k_2}} + \mu - U \right), 
\end{align}
with $\mathcal{F}(\mathbf{k_i + q}) = \frac{1}{2} \Theta\left( \epsilon_\mathbf{k_i + q} - \mu + U \right) \Theta \left( -\epsilon_\mathbf{k_i + q} + \mu \right) + \Theta\left( -\epsilon_\mathbf{k_i + q} + \mu -U  \right)$. Note that the Heaviside functions that appear in Eqs. \eqref{n0} to \eqref{n2} essentially define the momentum regions $\Omega_i$. These functions are depicted in Fig. \ref{Overlap}, which shows a possible overlap between  $\Omega_i$ for a particular choice of pair momentum $\mathbf{q}$ and chemical potential $\mu$.

We will consider the case in which $\mathbf{q=Q}$ [where $\mathbf{Q}=(\pi,\pi)$]. This choice is related to the fact that it has been recently demonstrated for different correlated models via both analytical and numerical methods  \cite{zhu2024exact,liu2024enhanced,wang2024pair,yue2024} that the PDW susceptibility at $\mathbf{q=Q}$ indicates that this phase turns out to be a close competitor to SC. {We also note that because of this particular finite pair momentum, one does not need to consider $-\mathbf{Q}$ in order to produce the PDW state.} This PDW order emerges, e.g., in fundamental strongly correlated models such as in the two-dimensional $\sigma_z$-Hubbard model \cite{zhu2024exact}, the Hubbard model defined on a bilayer square lattice \cite{liu2024enhanced} and the $t-J$ model on a square lattice \cite{yue2024}. {Last, we mention that the PDW at $\mathbf{Q}=(\pi,\pi)$ has close connections with the so-called $\eta$ pairs, which turn out to be exact (high-energy) eigenstates of the Hubbard model displaying off-diagonal long-range order \cite{CNYang_1989}.}

We now construct a PDW wavefunction $\ket{\psi_\mathbf{q}}$ from the pair creation operator $b^\dagger_{\mathbf{k+q}} = c^{\dagger}_{\mathbf{k+q},\uparrow} c^{\dagger}_{\mathbf{-k} \downarrow}$ by acting on the ground state of the HK model given by $\ket{g}$. Consequently, we have:
\begin{align}
    \ket{\psi_\mathbf{q}} = \left(  \sum_{\mathbf{k_0} } \alpha_{\mathbf{k_0}}\, b^\dagger_{\mathbf{k_0+q}} + \sum_{\mathbf{k_1}} \beta_\mathbf{k_1}\, b^\dagger_{\mathbf{k_1+q}} \right) \ket{g}.
\end{align}
The normalization of this state yields:
\begin{align}\label{NormalizationCondition}
\sum_{\mathbf{k_0} } \abs{\alpha_{\mathbf{k_0}}}^2 \left( 1 - n^\mathbf{q}_0(\mathbf{k_0}) \right) + \frac{1}{2}\sum_{\mathbf{k_1} } \abs{\beta_{\mathbf{k_1}}}^2 \left( 1-n^\mathbf{q}_1(\mathbf{k_1}) \right)=1.
\end{align}

To obtain the pair binding energy $E^{(b)}_\mathbf{q} \equiv E-E_0$ associated with this PDW state, which represents the energy gain of forming a Cooper pair with a center-of-mass momentum $\mathbf{q}$ in the model, we compute the following average energies $E_0=\bra{g} (H_{HK} + H_V) \ket{g}$ and $E=\bra{\psi_\mathbf{q}} (H_{HK} + H_V) \ket{\psi_\mathbf{q}}$. By setting $V_{\mathbf{q}}=V^{\prime}$, we find:
\begin{widetext}
\begin{equation}
\begin{aligned}\label{EBound}
    E^{(b)}_\mathbf{q} & = \sum_{\mathbf{k} \in \Omega_0} \left|\alpha_{\mathbf{k}}\right|^{2} [\xi_{\mathbf{k+q}} + \xi_{\mathbf{k}}] (1 - n^\mathbf{q}_{0}(\mathbf{k})) + \sum_{\mathbf{k} \in \Omega_1} \left|\beta_{\mathbf{k}}\right|^{2} [\xi_{\mathbf{k+q}} + \xi_{\mathbf{k}}] (1 - n^\mathbf{q}_{0}(\mathbf{k}))  + U \sum_{\mathbf{k} \in \Omega_0} \left| \alpha_{\mathbf{k}} \right|^2 (1 - n^\mathbf{q}_0(\mathbf{k})) n^\mathbf{q}_0(\mathbf{k}) \\
    &\quad + \frac{U}{2} \sum_{\mathbf{k} \in \Omega_1} \left| \beta_{\mathbf{k}} \right|^2 (1 - n^\mathbf{q}_0(\mathbf{k})) \left( \frac{1}{2} + n^\mathbf{q}_1(\mathbf{k}) \right) - V' \sum_{\mathbf{k}\in\Omega_{0},\mathbf{k}^{\prime}\in\Omega_{0}} \alpha_{\mathbf{k}}^* \alpha_{\mathbf{k'}} (1 - n^\mathbf{q}_{0}(\mathbf{k})) (1 - n^\mathbf{q}_{0}(\mathbf{k'})) \\
    &\quad - \frac{V'}{2} \sum_{\mathbf{k}\in\Omega_{0},\mathbf{k}^{\prime}\in\Omega_{1}} (\alpha_{\mathbf{k}}^* \beta_{\mathbf{k'}} + \alpha_{\mathbf{k}} \beta_{\mathbf{k'}}^*) (1 - n^\mathbf{q}_{0}(\mathbf{k})) (1 - n^\mathbf{q}_{1}(\mathbf{k'}))  - \frac{V'}{4} \sum_{\mathbf{k}\in\Omega_{1},\mathbf{k}^{\prime}\in\Omega_{1}} \beta^{*}_{\mathbf{k}} \beta_{\mathbf{k'}} (1 - n^\mathbf{q}_{1}(\mathbf{k})) (1 - n^\mathbf{q}_{1}(\mathbf{k'})).
\end{aligned}
\end{equation}
\end{widetext}

\begin{figure}[t]
     \centering
     \hspace{-0.55cm}
     \includegraphics[width=1.05\linewidth]{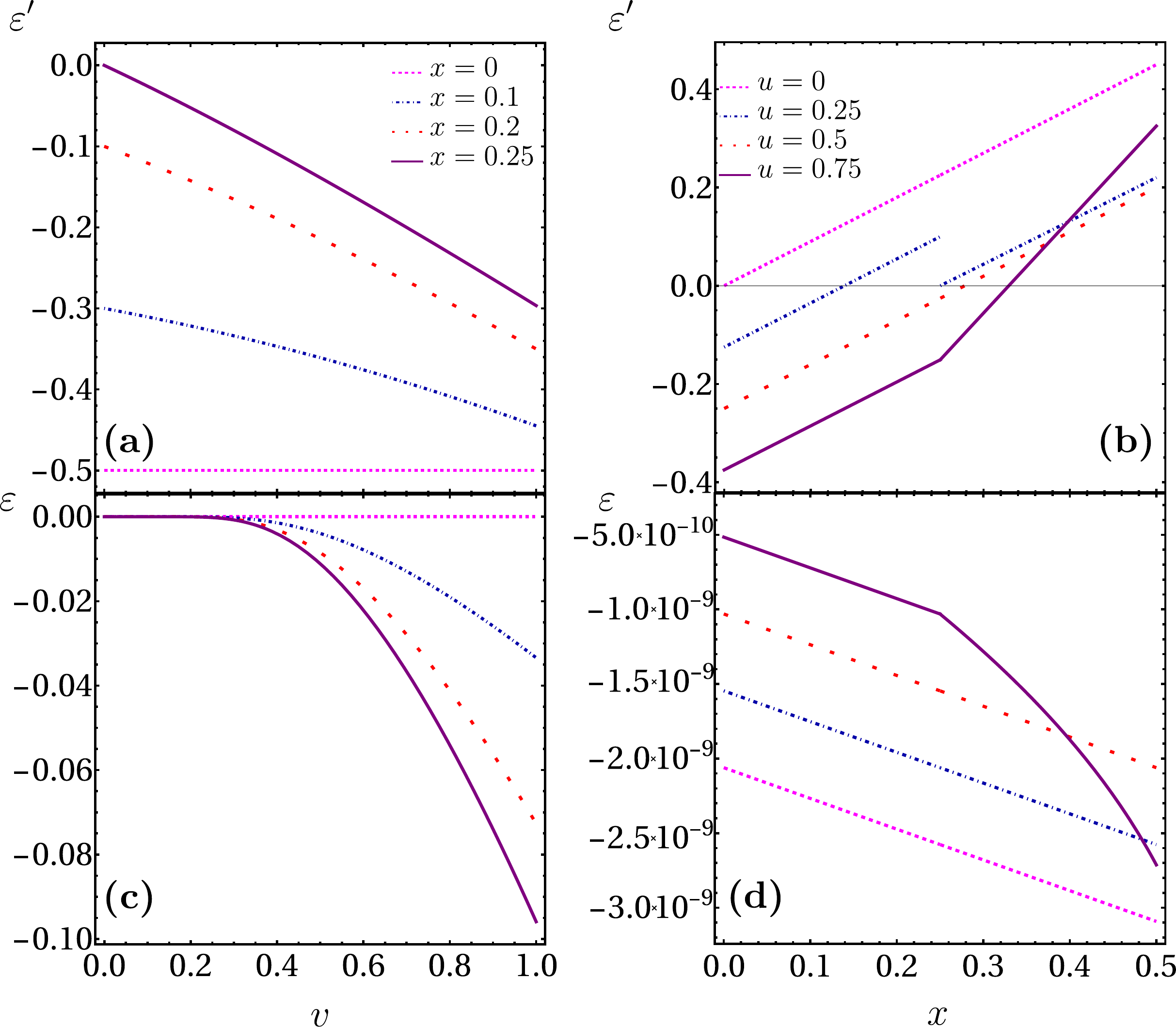}
     \caption{Cooper pair binding energy $\varepsilon'$ and $\varepsilon$ associated, respectively, with the emergence of PDW and SC in the {HK-SC} model: (a) $\varepsilon'$ and (c) $\varepsilon$ as a function of the interactions $v'=v$ for different doping parameters $x$ (we set $u=1$), and (b) $\varepsilon'$ and (d) $\varepsilon$ as a function of the doping parameter $x$ for different choices of the repulsion $u$ (we set $v'=v=0.1$).}
     \label{fig2}
\end{figure}

Next, we follow a standard variational approach \cite{Cooper_1956,Phillips2020,Li2022} by defining a function $\mathcal{Q}=E^{(b)}_\mathbf{q}-\lambda'(\langle{\psi_\mathbf{q}}|{\psi_\mathbf{q}}\rangle-1)$ subject to the constraint of the renormalization condition $\langle{\psi_\mathbf{q}}|{\psi_\mathbf{q}}\rangle=1$ given by Eq. \eqref{NormalizationCondition}, where $\lambda'=E^{(b)}_{\mathbf{q}}$ corresponds to a Lagrange multiplier. Employing the variational conditions given by $\partial \mathcal{Q}/\partial \alpha^*_{\mathbf{k}}=0$ and $\partial \mathcal{Q}/\partial \beta^*_{\mathbf{k}}=0$, the equations for the coefficients $\alpha_{\mathbf{k}}$ and $\beta_{\mathbf{k}}$ can be derived in a straightforward manner.
Summing $\alpha_{\mathbf{k}}$ over the momentum region $\Omega_0$ and $\beta_{\mathbf{k}}$ over the $\Omega_1$ region, we finally obtain the following self-consistent equation:
\begin{align}\nonumber
\frac{1}{V'} =& \sum_{\mathbf{k} \in \Omega _0} \frac{(1- n^{\mathbf{q}}_{0}(\mathbf{k}))}{\xi_{\mathbf{k + q}} + \xi_{\mathbf{k}} + U n^{\mathbf{q}}_{0}(\mathbf{k}) - \lambda'} \\\label{Self-Consistent-KSpace}
&+ \frac{1}{2} \sum_{\mathbf{k} \in \Omega _1} \frac{(1- n^{\mathbf{q}}_{1}(\mathbf{k}))}{\xi_{\mathbf{k + q}} + \xi_{\mathbf{k}} + U (n^{\mathbf{q}}_{1}(\mathbf{k})+\frac{1}{2}) - \lambda'}.
\end{align}
The above expression can be computed in the energy representation using the density of states (DOS) denoted by $\rho (\omega)$. For simplicity, we consider here a constant DOS given by $\rho(\omega) = \sum_{\mathbf{k}}  \delta(\omega - \xi_{\mathbf{k}})=1/W$ (for $-W/2\leq\omega\leq W/2$). Therefore, we have $\rho(\omega)=\rho_0(\omega)+\rho_1(\omega)+\rho_2(\omega)$, with $\rho_i(\omega)$ corresponding to the contribution of the momentum region $\Omega_i$. For the term associated with the region $\Omega_0$, we have:
\begin{align}\label{rho0}
\rho_{0}\left(\omega\right) = \sum_{\mathbf{k} \in \Omega_{0}} \delta\left(\omega - \xi_{\mathbf{k}}\right) = \Theta\left(\omega\right)\rho\left(\omega + \mu\right),
\end{align}
whereas, for $\mathbf{k} \in \Omega_1$, the corresponding quantity becomes:
\begin{align}\nonumber
\rho_{1}\left(\omega\right) =&   \frac{1}{2} \sum_{\mathbf{k} \in \Omega_{1}} \left[ \delta\left(\omega - \xi_{\mathbf{k}}\right) + \delta\left(\omega - \xi_{\mathbf{k}} - U  \right) \right]\\\nonumber
=& \frac{1}{2}\left[ \Theta \left( -\omega \right) \Theta\left( \omega + U \right) \rho\left(\omega + \mu\right) \right.\\\label{rho1}
&\left. +\Theta\left(-\omega + U\right)\Theta\left(\omega\right)\rho\left(\omega + \mu - U\right)\right],
\end{align}
and finally, if $\mathbf{k} \in \Omega_2$, then we obtain that
\begin{align}\label{rho2}
\rho_{2}\left(\omega\right) = \sum_{\mathbf{k} \in \Omega_{2}} \delta\left(\omega - \xi_{\mathbf{k}}\right) = \Theta(-\omega)\rho(\omega-\mu).
\end{align}

{Since} $\epsilon_{\mathbf{k+Q}} = -\epsilon_{\mathbf{k}}$, and inserting Eqs. \eqref{rho0} and \eqref{rho1} into the self-consistent equation given by Eq. \eqref{Self-Consistent-KSpace}, we get:
\begin{widetext}
\begin{align}\label{Self-Consistent-ESpace}
    \frac{1}{v'} &= \int_{0}^{\frac{1}{2}-\widetilde{\mu}} \frac{(1- n_{0}(\omega))d\omega}{-2\widetilde{\mu} + {u}\, n_{0}(\omega) - \lambda'} \Theta(\omega) \Theta\left(-\omega - \tilde{\mu} + \frac{1}{2}\right) \Theta\left(\omega + \tilde{\mu} + \frac{1}{2}\right) \nonumber\\
    &+ \frac{1}{4} \int_{-{u}}^{0} \frac{(1- n_{1}(\omega))d\omega}{-2\widetilde{\mu} +{u}(n_{1}(\omega)+\frac{1}{2}) - \lambda'} \Theta(-\omega) \Theta(\omega+{u}) \Theta \left(-\omega -\tilde{\mu} + \frac{1}{2} \right)\Theta \left(\omega +\tilde{\mu} + \frac{1}{2} \right)  \nonumber\\
    & + \frac{1}{4} \int_{0}^{{u}} \frac{(1- n_{1}(\omega))d\omega}{-2\widetilde{\mu} + {u} (n_{1}(\omega)+\frac{1}{2}) - \lambda'} \Theta(\omega) \Theta(-\omega + {u}) \Theta \left( \omega + \tilde{\mu}-u + \frac{1}{2} \right) \Theta \left(-\omega - \tilde{\mu}+u + \frac{1}{2} \right),
\end{align}
\end{widetext}
where we have also defined the dimensionless variables $v'=V'/W$ and $\varepsilon'=\lambda'/W$. By integrating Eq. \eqref{Self-Consistent-ESpace}, we obtain the (rescaled) pair-binding energy $\varepsilon'$ associated with the emergence of the PDW phase in the {HK-SC} model.

In Fig. \ref{fig2}(a), we plot the resulting Cooper pair binding energy $\varepsilon'$ as a function of the pair interaction $v'$, for several doping parameters. We observe from this figure that in the underdoped regime up to a critical doping ($x_c\lesssim 0.25)$, $\varepsilon'$ remains negative for any $v'$. This means that this pairing channel is capable in principle of generating an instability towards the formation of a PDW in the model for this doping range. Moreover, in
Fig. \ref{fig2}(b), we also display the same quantity $\varepsilon'$ obtained from the self-consistent equation in Eq. \eqref{Self-Consistent-ESpace}, but now as a function of the doping parameter $x$ for several choices of the dimensionless interaction $u$. As a result, we observe that, while increasing the interaction $U$ (for $U\leq W)$ is beneficial for the emergence of PDW order in the model with $\varepsilon'$ becoming more negative, further doping away from half-filling turns out to be instead detrimental to this phase. The latter behavior is confirmed by the fact that the pair-binding energy $\varepsilon'$ in
Fig. \ref{fig2}(b) is positive on sufficient doping, indicating that forming a Cooper pair with finite center-of-mass momentum becomes unfavorable in the model for higher dopings.

To investigate the competition between the tendency towards the formation of a PDW and the familiar SC order in the {HK-SC} model, we also revisit the pair binding energy associated with a pairing state with $\mathbf{q}=0$. By defining $V_{\mathbf{q}=0}=V$, we find the self-consistent equation
\begin{equation}\label{BCS_inst}
    \frac{1}{V}=\sum_{\mathbf{k}\in\Omega_0}\frac{1}{2\xi_\mathbf{k}-\lambda}+\frac{1}{4}\sum_{\mathbf{k}\in\Omega_1}\frac{1}{2\xi_\mathbf{k}+U-\lambda},
\end{equation}
where $\lambda=E^{(b)}_{\mathbf{q}=0} $. By solving Eq. \eqref{BCS_inst} in the limit of $|E^{(b)}_{\mathbf{q}=0}|< U$ and $V\ll W$, we get the results plotted in Figs. \ref{fig2}(c) and \ref{fig2}(d),
where $\varepsilon=\lambda/W$ and $v=V/W$. 

As shown in Fig. \ref{fig2}(c), we obtain that a small nonzero $v$ is sufficient for a SC phase to emerge in the model. This is related to the fact that the pair binding energy $\varepsilon$ associated with the SC instability decays exponentially as a function of $1/v$. This agrees with the original BCS result \cite{Cooper_1956,BCS_1957}, and it is valid in the present model for any interaction $U>0$. We also point out that our results are in good agreement with the conclusions of Ref. \cite{Phillips2020}, where a similar expression was derived for $U<W$ at half-filling. However, it agrees only partially with the result of Ref. \cite{Li2022}, where a slightly different formula for the pair binding energy $E^{(b)}_{\mathbf{q}=0}$ was obtained. The latter disagreement stems from a different prefactor multiplying $\xi_{\mathbf{k}}$ in the term of Eq. \eqref{BCS_inst} corresponding to the sum over $\Omega_1$ that we derived here compared with the result in Ref. \cite{Li2022}. In Fig. \ref{fig2}(d), we display our results of $\varepsilon$ as a function of the doping parameter $x$ for several interactions $u$. We note that although $\varepsilon$ always remains negative, its absolute value decreases by raising $U$, indicating that increasing the $U$ interaction is actually worse for SC (unlike the case for PDW order). Moreover, doping away from half-filling in the model has a positive effect on the SC phase, since it enhances the tendency toward its formation.

\begin{figure}[t]
\centering
\centering \includegraphics[width=1\linewidth]{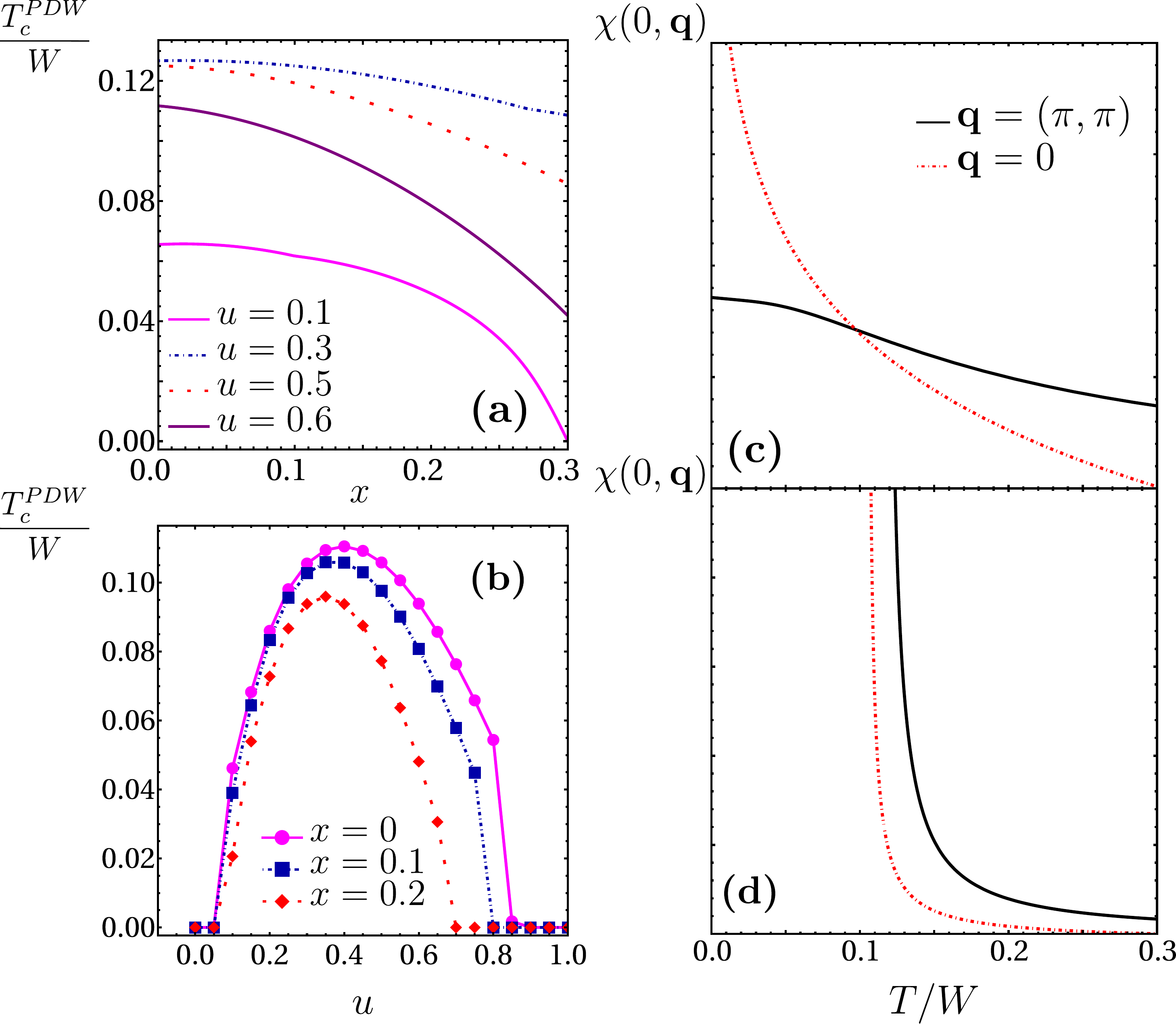}
\caption{(a) Critical temperature $T_c^{PDW}$ in the {HK-SC} model as a function of the doping parameter $x$ (for some choices of $u$) for $v'=0.75$. (b) Critical temperature $T_c^{PDW}$ as a function of $u$ for some doping parameters $x$. [(c) and (d)] Static pair susceptibilities $\chi(0,\mathbf{q})$ for SC and PDW as a function of $T$ for $v=v'\rightarrow 0$ and $v=v'=0.75$, respectively.}\label{Plot}
\end{figure}

\section{Critical temperatures associated with PDW and SC orders} \label{Tc}

The previous result regarding the calculation of the Cooper-pair binding energy $E^{(b)}_{\mathbf{q}}$ associated with PDW order is suggestive of an important tendency towards the formation of such a phase in the {HK-SC} model. However, in order to establish the final fate of the PDW phase (and also of SC order), we must compute the pair susceptibility as a function of temperature, which is given by
\begin{equation}\label{exact_susc}
\chi(i\nu_n,\mathbf{q})=\int_{0}^{\beta}d\tau\, e^{i\nu_n\tau}\langle T\Delta_{\mathbf{q}} (\tau)\Delta_{\mathbf{q}}^{\dagger}\rangle_{V_{\mathbf{q}}},
\end{equation}
where $\Delta_{\mathbf{q}}=\sum_{\mathbf{k}}b_{\mathbf{k+q}}=\sum_{\mathbf{k}}c_{\mathbf{-k} \downarrow}c_{\mathbf{k+q},\uparrow} $, $\langle ...\rangle_{V_{\mathbf{q}}}$ refers to the expectation value in the grand-canonical ensemble with density operator $e^{-\beta H}/Z$, with $Z$ being the partition function, $\beta=1/k_B T$ (henceforth, we will use natural units such that $k_B=1$), and $\nu_n=2\pi n/\beta$ is a bosonic Matsubara frequency. 

We will perform here an exact calculation of Eq. \eqref{exact_susc} in an intermediate-to-low temperature regime, i.e., such that $T\ll U,W$. To this end, we will follow the demonstration of Ref. \cite{Phillips2020} where $\chi(i\nu_n,\mathbf{q})$ of the HK model in the presence of a finite $V_{\mathbf{q}}$ is related to the `bare' susceptibility $\chi_0(i\nu_n,\mathbf{q})$ for zero pairing interactions (i.e., $V_{\mathbf{q}}=0$) by means of the Dyson equation:
\begin{equation}
\chi=\chi_0+V_{\mathbf{q}}\chi_0\chi.
\end{equation}
The ratio $L\equiv -V_{\mathbf{q}}\,\chi/\chi_0$ then satisfies 
\begin{equation}
L=\frac{1}{\chi_0-1/V_{\mathbf{q}}},
\end{equation}
and, consequently,
the critical temperatures $T_c^{PDW}$ and $T_c^{SC}$ associated with an instability of PDW and SC in the model can be obtained from the following expressions:
\begin{eqnarray}
&&\chi_0(i\nu_n=0,\mathbf{q}=0)\bigg|_{T=T_c^{SC}}=1/V,\\
&&\chi_0(i\nu_n=0,\mathbf{q}=\mathbf{Q})\bigg|_{T=T_c^{PDW}}=1/V'.
\end{eqnarray}
By using Eq. \eqref{ExactGF}, we can show that the above calculation of the susceptibilities can be divided \cite{Phillips2020} into contributions from the lower Hubbard band (with energy dispersion $\xi^{l}_{\mathbf{k}} = \xi_{\mathbf{k}}$ and number operator $n_{\mathbf{k}, \sigma}^{l} = 1 - \langle n_{\mathbf{k}, \bar{\sigma}} \rangle$) and the upper Hubbard band (with energy dispersion $\xi_{\mathbf{k}}^{u} = \xi_{\mathbf{k}} + U$ and number operator $n_{\mathbf{k},\sigma}^u = \langle n_{\mathbf{k}, \bar{\sigma}} \rangle$), yielding:
\begin{align}\label{Susc_ab}
    \chi_0(i\nu_n,\mathbf{q}) =& \sum_{\mathbf{k},a,b} n^{a}_{\mathbf{k},\sigma} n^{b}_{\mathbf{k},\sigma} \frac{f\bigl( \xi^{a}_{\mathbf{k+q}} \bigl) + f\bigl( \xi^{b}_{-\mathbf{k}} \bigl)-1 }{i\nu_n-\xi^{a}_{\mathbf{k+q}} - \xi^{b}_{-\mathbf{k}}},
\end{align}
where $a,b = u \mbox{ or } l$ and $f(\xi)$ is the Fermi-Dirac distribution function.

In the SC channel ($\mathbf{q}=0$), the only finite contributions for $T\ll U$ arise from the same Hubbard bands (i.e., for $a=b$). Consequently, we arrive at the result:
\begin{align}\label{Susc_SC}
    \chi_0(0,\mathbf{q}=0) =& \int_{-W/2}^{W/2} d \omega\, \rho'(\omega) \frac{\tanh \left( \frac{\beta \omega}{2} \right)}{2 \omega},
\end{align}
where $\rho'(\omega) = \rho_0(\omega) + (1/2) \rho_{1}(\omega) + \rho_{2} (\omega)$ is an effective density of states. Following standard methods (see, e.g., Ref. \cite{Phillips2020}), we integrate exactly the above equation for any chemical potential $\mu$. In the following, we obtain two results depending on whether the momentum region $\Omega_2$ appears in the band or not. For the first case (i.e., $-W/2 \leq \mu_c-U \leq W/2$), the SC critical temperature is given by:
\begin{align}\label{Tc_SC1}
    T_c^{SC} &=\frac{2 e^{\gamma}}{\pi} U^{1/5} \left( \frac{W}{2} - U + \mu_c \right)^{2/5}\left(\frac{W}{2} - \mu_c \right)^{2/5} e^{-\frac{4W}{5V}},
\end{align}
where $\gamma\approx 0.577$ is the Euler-Mascheroni constant. As for the second case (i.e., $\mu_c  <U - W/2$), we obtain that:
\begin{align}\label{Tc_SC2}
    T_c^{SC} &=\frac{2 e^{\gamma}}{\pi} U^{1/5} \left( \frac{W}{2}-\mu_c \right)^{4/5} \left( \frac{W+2\mu_c}{2U-W-2\mu_c} \right)^{1/5} e^{-\frac{8W}{5V}}.
\end{align}
We point out that $\mu_c$ in Eqs. \eqref{Tc_SC1} and \eqref{Tc_SC2} refers to the chemical potential at the SC critical temperature\footnote{We also note that the regime $\mu_c > U+W/2$ is not discussed here, because this would correspond to a strongly electron-doped situation in the model, which does not yield any new physics.}. Interestingly, for $U<W$ at half-filling ($\mu_c=U/2$), there is indeed a SC instability at a finite critical temperature in the model, in good agreement with the conclusions of Ref. \cite{Phillips2020}. By contrast, for $U>W$ at the same doping regime ($\mu_c=W/2$), there is not such a pairing instability (i.e., $T_c^{SC}=0$) due to the presence of the Mott gap in the latter regime.

Regarding the calculation of the critical temperature associated with PDW with $\mathbf{Q} = (\pi,\pi)$ in the model, we point out that in addition to the $ll$ and $uu$ contributions in Eq. \eqref{Susc_ab} coming from the lower and upper Hubbard bands, the mixed terms (i.e., $ul$ and $lu$) also contribute to the PDW channel. By making a similar analysis as in the SC case, we arrive at
\begin{align}\nonumber
   \chi_0(0,\mathbf{Q}) =& \sum_{\mathbf{k},a,b}  \frac{n^{a}_\mathbf{k+Q} n^{b}_\mathbf{k}}{\xi^a_{\mathbf{k+Q}}+\xi^b_{\mathbf{k}}}\\ 
   & \times \frac{\sinh \left( \frac{\beta \omega_{\mathbf{k,Q}}^{ab} } {2} \right)}{\cosh \left( \frac{\beta \omega_{\mathbf{k,Q}}^{ab}}{2}  \right) + \cosh \left( \frac{\beta}{2} \left( \xi^{a}_{\mathbf{k+Q}} - \xi^b_{\mathbf{k}} \right) \right)},\label{PDW_equation}
\end{align}
where $a,b = u \mbox{ or } l$ and $\omega_{\mathbf{k,Q}}^{ab}= \xi^a_{\mathbf{k+Q}} + \xi^b_{\mathbf{k}} $. Since $\xi_{\mathbf{k+Q}}= -\xi_{\mathbf{k}}-2\mu$, $\omega^{ab}_{\mathbf{k,Q}}$ becomes independent of $\mathbf{k}$. This means that the integral in Eq. \eqref{PDW_equation} can also be performed exactly.
The analytical result concerning this calculation turns out to be too cumbersome and, therefore, we leave it to Appendix.

In Fig. \ref{Plot}(a), we plot the dependence of $T_c^{PDW}$ as a function of $x$ (for different values of $u$). In this plot, we observe that $T_c^{PDW}$ {exhibits a nonmonotonic behavior as a function of $u$}.
It gets strongly suppressed with doping for small $u$, but this reduction becomes relatively milder for intermediate $u$, which indicates that the PDW phase is indeed more robust for the latter interaction regime. In Fig. \ref{Plot}(b), we depict the behavior of $T_c^{PDW}$ as a function of $u$ (for different doping parameters $x$), {which provides further evidence of the aforementioned nonmonotonic behavior}. While we confirm that doping is detrimental to PDW, we observe that there are minimum and maximum values for the (rescaled) HK interaction $u$, where a PDW phase can in principle appear at finite temperatures in the {HK-SC} model. Furthermore, we obtain for a finite $v'$ that the PDW susceptibility wins against SC for low doping (provided that $u$ is larger than a critical value), while it loses to SC at sufficiently large dopings (for any $u$). These results are displayed in Figs. \ref{Plot}(c) and \ref{Plot}(d). In this connection, we point out that a minimal $v'$ is necessary for a long-range PDW order to emerge in the present model at $T>0$. If $v'$ turns out to be smaller than such a critical value (whose magnitude decreases as a function of $u$ and increases as a function of $x$), then the PDW susceptibility at $\mathbf{Q} = (\pi,\pi)$, although strongly enhanced with lowering $T$, remains nevertheless nondivergent at finite temperatures [see Fig. \ref{Plot}(c)]. This result implies that for $v'\rightarrow 0$ one cannot interpret the emergence of a PDW order as a true instability in the HK model, but rather as a correlated phase that exhibits prominent PDW fluctuations at finite temperatures that emerge on doping a Mott insulator. 

\begin{figure}[t]
\centering
\centering \includegraphics[width=0.7\linewidth]{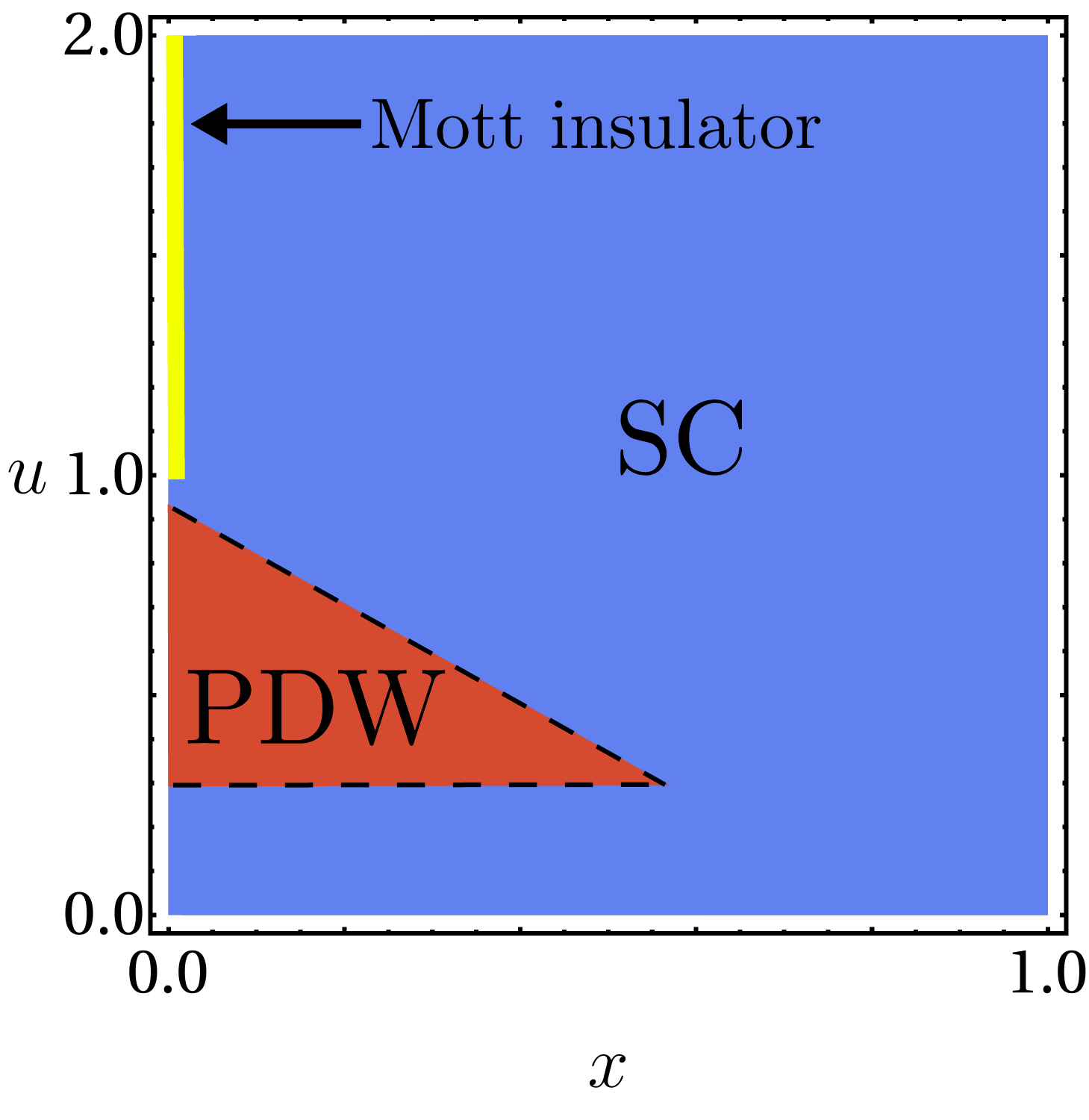}
\caption{Schematic phase diagram displaying the {leading} pairing {instabilities} (PDW or SC) as a function of the interaction $u$ and doping parameter $x$ in the {HK-SC} model with {two} pairing interactions $v$ and $v'$. The dashed line is only a guide to the eye. Here, we set $v=v'=0.75$. For completeness, we also depict the Mott insulating phase in the model for $u>1$ and $x=0$. }\label{Diagram}
\end{figure}

A schematic phase diagram displaying {the leading pairing instabilities} in the {HK-SC} model as a function of both $u$ and $x$ is depicted in Fig. \ref{Diagram}. 
From this plot, we observe that a sufficiently weak $U$-coupling is always beneficial to SC order. However, the tendency towards the formation of a PDW at $\mathbf{Q}=(\pi,\pi)$ becomes more favorable within an intermediate $u$ regime and low doping in the phase diagram, for a moderate $v'$. We also note that stronger interactions (i.e., $u>1)$ in the {HK-SC} model become in fact detrimental to both SC and PDW phases, with the effect on the former being less severe than in the latter phase. Moreover, far from half-filling, the SC instability always tends to dominate. We point out that all these results are in agreement with our calculation in the previous section regarding the analysis of the pair binding energies $E^{(b)}_{\mathbf{q}}$ associated with SC and PDW phases and, for this reason, the latter provides yet another confirmation of the scenario obtained in this work. We thus conclude from our data that low dopings up to a critical doping regime and intermediate $U$ interactions turn out to be crucial for a PDW fluctuating regime to appear at finite temperatures in the {HK-SC} model. Interestingly, our results share some qualitative similarities with the conjecture of the possible existence of strong PDW fluctuations at finite temperatures in underdoped cuprates (well-known to be the archetypal doped Mott insulator), which have been argued to be relevant in order to understand the pseudogap phase displayed by those compounds \cite{PALee_2014,Agterberg_2020}. It also agrees with the fact that superconductivity emerges from such a PDW fluctuating regime with further doping.

\section{Conclusions and Outlook}\label{Conclusions}

The HK model \cite{HK_model} is an interesting minimal correlated model that describes both a Mott insulator and a non-Fermi liquid with an instability towards a SC phase \cite{Phillips2020}, which turns out to be analytically solvable for any value of the interaction $U$ and doping parameter $x$. For this reason, we believe this model may open a new perspective in the study of strongly correlated systems. In the present work, we have shown that its phase diagram is in fact richer, by demonstrating via an exact analysis of the pairing susceptibilities of the model that it also describes the appearance of a PDW fluctuating phase [with wavevector $\mathbf{Q}=(\pi,\pi)$] at finite temperatures for intermediate values of the $U$ interaction and within a low doping up to a critical doping regime. In other words, PDW indeed emerges as a close competitor to SC within these interaction and doping regimes in the {HK-SC} model. With further doping, we recover the expected SC phase with zero pair momentum. 

Recently, it has been argued \cite{Huang_FixedPoint} that the HK model should correspond to a stable fixed point in the renormalization group flow of the paradigmatic and hard-to-solve Hubbard model, e.g., in two dimensions, which is widely assumed to be a central model in order to describe many strongly correlated materials. For this reason, one could in principle expect some degree of universality in the predictions coming from the exact solubility of the HK model, and try to use these results to interpret some qualitative aspects of the physics of correlated materials.

Furthermore, in order to achieve detailed quantitative agreement with realistic systems, a new avenue of research has been put forward recently in terms of an orbital extension of this model, dubbed orbital HK model \cite{BBradlyn,Orbital_HK_1, Orbital_HK_2}. Indeed, it has been shown in Ref. \cite{Orbital_HK_1} that by decorating each $\mathbf{k}$ state in the model by $n$ local (orbital) degrees of freedom, one can achieve rapid numerical convergence as $n$ increases, paving the way for a new computational tool to address strongly correlated systems. In this connection, quantitative agreement was obtained, e.g., in the one-dimensional Hubbard model reproducing the results obtained by exact diagonalization and DMRG approach \cite{Orbital_HK_1}, and in the Mott insulating phase of the two-dimensional Hubbard model that emerges regardless of the strength of the local interaction $U$ at half-filling \cite{Orbital_HK_1}, which is consistent with quantum Monte Carlo simulations.

Another promising direction of investigation regarding the HK model concerns the calculation of various transport properties of the non-Fermi liquid phase that appears in this system. Recent theoretical work \cite{Sangiovanni_HK} has argued that the Kubo formalism is perhaps not a good strategy to compute these properties in the model since by using this approach one obtains some inconsistencies that originate from the presence of a local interaction in momentum space. However, a suitable framework to tackle non-Fermi liquids in general turns out to be the memory matrix approach \cite{forster-book,Patel-PRB,Hartnoll_PRB_2014,Freire-AP_2017,Berg_memory_mat,Freire-AP_2020,ips_hermann1,Pangburn_2023,Hermann_review}, which employs a coarse-graining procedure of separating the irrelevant modes and deriving an effective theory for the relevant ones by means of projecting the transport theory onto a basis of 
nearly conserved operators. Therefore, we believe that by solving this model using the latter methodology, one could in principle address the important question of whether strange metal physics appears in the HK model or not.  

\section*{Acknowledgment}  

H.F. acknowledges funding from the Conselho Nacional de Desenvolvimento Cient\'{i}fico e Tecnol\'{o}gico (CNPq) under grant numbers 311428/2021-5 and 404274/2023-4. 

\appendix

\section*{Appendix: Results for $T_c^{PDW}$ and $\mu(T)$}\label{Appendix_A}

The susceptibility for PDW order displayed in Eq. \eqref{PDW_equation} can be analytically calculated for the two regimes defined in the main text. In this Appendix, we show the corresponding expressions for completeness. For the first regime, in which $- W/2 \leq \mu'_c - U \leq W/2$, we get:
\begin{widetext}
\begin{align}\nonumber
    \chi_0 (0, \mathbf{Q}) =& \frac{1}{16 W \mu_c'\beta(U-2\mu_c')(U-\mu_c')} \Bigg[8 ({U} - \mu_c' )  \Big( {U}\beta ( {U} - 2\mu_c' ) + {U} \ln ( 2 (1 + {e}^{\beta{U}} ) ) + 4\mu_c'\Big[ \ln ( {e}^{\frac{\beta{U}}{2}} + {e}^{\frac{1}{2}\beta (W + {U- 2\mu_c'} ) } ) \\\nonumber
     &- \ln ( {e}^{\beta/2} + {e}^{\beta ( {U} - \mu_c' ) } )\Big]  
    - {U} \Big[ \ln ( 1 + {e}^{\beta ( {U} - 2\mu_c' ) } ) + \ln ( {e}^{\beta{U}} ( 1 + {e}^{-2\beta\mu_c'} ) ) \Big] \Big)\nonumber\\
    &+ U \left\{ -\beta ( U - 2\mu_c' ) ( 5U - \mu_c' ) + ( \mu_c' - U ) \ln ( 1 + {e}^{2\beta\mu_c'} ) - 4U \ln ( 2 ) + \mu_c' \ln ( 16 ) - 4\mu_c' \ln ( 1 + {e}^{\beta U} ) \right.\nonumber\\\label{A1}
    &\left. + 4\mu_c' \ln ( 1 + {e}^{\beta ( U - 2\mu_c' ) } ) + 5 ( U - \mu_c' ) \ln ( {e}^{U\beta} ( 1 + {e}^{-2\beta\mu_c'} ) ) \right\} \Theta ( \mu_c' ) \Bigg].
\end{align}
As for the second regime, such that $\mu'_c < U- W/2$, we obtain:
\begin{align}\label{A2}
     \chi_0 (0, \mathbf{Q}) =& \frac{\ln \left( \cosh \left( \frac{\beta U}{2} \right) \sech \left( \frac{\beta (U - 2 \mu_c')}{2} \right) \right) \Theta\left( \mu_c' \right)}{8 W \beta( U - \mu_c')} - \frac{ \ln \left( \cosh \left( \beta \mu_c' \right) - 
     \sinh \left( \beta \mu_c' \right) \tanh^{-1} \left( \frac{\beta U }{2} \right) \right)^2 \Theta \left(\mu_c' \right)}{
    16 W \beta (U - \mu_c')}\nonumber \\
    &+ \frac{ \left( 2 \beta \mu_c' + 
    \ln \left( \frac{(1 + e^{\beta U}) (1 - \tanh \left( \beta \mu_c' \right)}{ e^{\beta U} + e^{2 \beta \mu_c'}} \right)  \right) \Theta \left(\mu_c' \right)}{4 W \beta (U - 2 \mu_c')} - \frac{\ln \left( \cosh \left( \frac{\beta (W - 2 \mu_c')}{4} \right) \sech \left(\beta \mu_c'\right) \sech
     \left( \frac{\beta (W + 2 \mu_c')}{2} \right) \right)}{4 W \beta \mu_c'}\nonumber \\
     &+ \frac{\beta \mu_c' + \ln \left( \frac{2}{1 + e^{2 \beta \mu_c'}} \right)}{4 W \beta \mu_c' } \Theta \left( \mu_c' \right)
     - \frac{-3 \beta \mu_c' + 3 \ln \left( 1 + e^{2 \beta \mu_c'} \right) + 
     \ln \left( \frac{1}{8} \cosh \left( \frac{\beta (W - 2 \mu_c')}{4} \right) \sech
     \left( \frac{\beta (W + 2 \mu_c') }{4} \right) \right)}{4 W \beta \mu_c'} \nonumber\\
     &-\frac{ \left( 2 \beta \mu_c' + 2\ln \left( \frac{4}{ \left( 1 + e^{2 \beta \mu_c'} \right)} \right) \right) \Theta \left( \mu_c' \right)}{4 W \beta \mu_c'} + \frac{W \beta + 2 \beta \mu_c' - 2 \ln\left( \frac{e^{\frac{W \beta}{2}} \left( 1 + e^{2 \beta \mu_c'} \right)}{e^{\frac{W \beta}{2}} + e^{\beta \mu_c'}} \right) - 2 \ln \left(\frac{(e^{\beta U} + e^{\frac{\beta W}{2} + \beta \mu_c'}}{1 + e^{\beta U}} \right)}{8 W \beta (U - 2 \mu_c')} \nonumber\\
     &+ \frac{ \ln \left( \frac{(1 + e^{\beta (U - 2 \mu_c')}) (1 + e^{2 \beta \mu_c'})}{2 (1 + e^{\beta U}}) \right) \Theta \left(2 \mu_c' \right)}{8 W \beta (U - 2 \mu_c')} + \frac{ W \beta + 2 \beta \mu_c' - 2 \ln \left( \frac{1 + e^{2 \beta \mu_c'}}{1 + e^{\beta (-\frac{W}{2} + \mu_c')}} \right) - 2 \ln\left( \frac{e^{\beta U} + e^{\beta \left( \frac{W}{2} +\mu_c' \right)}}{1 + e^{\beta U}} \right)}{8 W \beta (U - 2 \mu_c')}\nonumber\\
     &+ \frac{ \ln\left( \frac{(1 + e^{\beta (U - 2 \mu_c')}) (1 + e^{2 \beta \mu_c'})}{2 (1 + e^{\beta U})} \right) \Theta \left( 2 \mu_c' \right)}{8 W \beta (U - 2 \mu_c'))} - \frac{ \left(2 \ln \left( \frac{2}{1 + e^{2 \beta \mu_c'}} \right) + 2 \ln\left( \frac{1 + e^{\beta U} }{e^{\beta U} + e^{2 \beta \mu_c'}} \right) \right) \Theta \left( \mu_c' \right)}{4 W \beta (U - 2 \mu_c')} \nonumber\\
     & - \frac{2 \bigg( -\beta \mu_c' + \ln \left( \frac{\left(1 + e^{2 \beta \mu_c'} \right) \left( e^{\beta U} + e^{\beta \left( \frac{W}{2} + \mu_c' \right)} \right)}{ \left( 1 + e^{\beta U} \right) \left( e^{ \frac{\beta W}{2}} + e^{\beta \mu_c'} \right)} \right) + 2 \beta \mu_c' \Theta \left(2 \mu_c' \right) \bigg)}{4 W \beta (U - 2 \mu_c')}.
\end{align}
\end{widetext}

We note that $\mu'_c$ in Eqs. \eqref{A1} and \eqref{A2} refers to the chemical potential at the PDW critical temperature. From the above expressions, we obtain the results related to the discussion of the PDW fluctuating phase that emerges at finite temperatures in the {HK-SC} model. The corresponding analysis appears in the main part of the text. Last, we show the dependence of the chemical potential as a function of $\beta$, $x$, and $U$ for the two regimes described above and in the main text. For the first regime, i.e., $- W/2 \leq \mu - U \leq W/2$, we have:
\begin{widetext}
\begin{align}
    \mu  =& \frac{1}{\beta} \ln \left(\frac{e^{\beta(U - W)} \left(-e^{\frac{3 \beta W }{2}} \left(e^{\beta W x} -1\right) + \sqrt{2} \sqrt{e^{\beta \left(-U + W (3 + x) \right)} \left( -e^{\beta U} + \cosh{\left( \beta W \right)} + \left(-1 + e^{ \beta U \beta} \right) \cosh\left(\beta W x\right) \right)}\right)}{\left( e^{\beta W (1 + x)} - 1\right)} \right),
\end{align}
whereas for the second regime, i.e., $\mu < U- W/2$, we get:
\begin{align}\label{A2}
    \mu  =& \frac{1}{\beta} \ln\left( \frac{e^{\frac{\beta}{2} \left( W+2U \right)}-2e^{\beta \left( W \left( \frac{1}{2} + x \right) + U \right)} +4 \sqrt{4 \left( e^{\beta W (1+x)} -1 \right) \left( e^{\beta (W+U)} - e^{\beta(Wx + U)} \right) + \left( e^{\frac{\beta}{2} \left( W + 2U \right)} - 2e^{\beta \left( W \left( \frac{1}{2} + x \right) + U \right)} \right)} }{2 \left( e^{\beta W \left( 1+x \right)} -1 \right)} \right) .
\end{align}
\end{widetext}


%

\end{document}